\documentclass[pra,twocolumn,amsmath,amssymb,floatfix,reprint,footinbib,superscriptaddress,longbibliography,showkeys]{revtex4-1}
\usepackage{dcolumn}
\usepackage{graphicx}
\usepackage{mathrsfs}
\usepackage{mdwlist}
\usepackage{subfigure}
\usepackage{booktabs}
\usepackage{multirow}
\usepackage{amsmath}
\usepackage{textcomp}
\usepackage{upgreek}
\usepackage{dsfont}
\usepackage{amstext}
\usepackage{amssymb}
\usepackage{amsbsy}
\usepackage{appendix}
\usepackage{soul}
\usepackage{threeparttable}
\usepackage{bbm}
\usepackage{amsthm}
\usepackage{graphicx}
\usepackage{lipsum}
\usepackage{xcolor}
\usepackage[none]{hyphenat}
\usepackage{textcomp}
\usepackage{color}
\usepackage[colorlinks,citecolor=blue]{hyperref}
\definecolor{Dgreen}{RGB}{0, 100, 0}
\usepackage{url}
\usepackage[colorlinks]{hyperref}
\hypersetup{%
	plainpages=true,
	breaklinks=true,  
	hypertexnames=false,  
	pageanchor=true,
	colorlinks=true,
	linkcolor={blue},
	citecolor={red},
	urlcolor={blue},
	anchorcolor={black}
}

\begin{document}

\title{Preparation of high-fidelity entangled cat states with composite pulses}
\author{Ge-Ge Gu}
\affiliation{Fujian Key Laboratory of Quantum Information and Quantum Optics, Fuzhou University, Fuzhou 350108, China}
\affiliation{Department of Physics, Fuzhou University, Fuzhou 350108, China}

\author{Dong-Sheng Li}
\affiliation{Fujian Key Laboratory of Quantum Information and Quantum Optics, Fuzhou University, Fuzhou 350108, China}
\affiliation{Department of Physics, Fuzhou University, Fuzhou 350108, China}

\author{Ye-Hong Chen}\thanks{yehong.chen@fzu.edu.cn}
\affiliation{Fujian Key Laboratory of Quantum Information and Quantum Optics, Fuzhou University, Fuzhou 350108, China}
\affiliation{Department of Physics, Fuzhou University, Fuzhou 350108, China}
\affiliation{Theoretical Quantum Physics Laboratory, Cluster for Pioneering Research, RIKEN, Wako-shi, Saitama 351-0198, Japan}

\author{Bi-Hua Huang}\thanks{hbh@fzu.edu.cn}
\affiliation{Fujian Key Laboratory of Quantum Information and Quantum Optics, Fuzhou University, Fuzhou 350108, China}
\affiliation{Department of Physics, Fuzhou University, Fuzhou 350108, China}

\author{Yan Xia}\thanks{xia-208@163.com}
\affiliation{Fujian Key Laboratory of Quantum Information and Quantum Optics, Fuzhou University, Fuzhou 350108, China}
\affiliation{Department of Physics, Fuzhou University, Fuzhou 350108, China}

\begin{abstract}
We propose a protocol for the preparation of high-fidelity entangled cat states with composite pulses. The physical model contains two Kerr-nonlinear resonators and a cavity. By properly designing the parameters, each Kerr-nonlinear resonator is confined in the cat-state subspace and the entangled cat states can be generated efficiently. We introduce composite two-photon drives with multiple amplitudes and frequencies to improve the fidelity of the entangled cat states in the presence of parameter errors. The performance of the protocol is estimated by taking into account the parametric errors and decoherence. Numerical simulation results show that, the protocol is robustness to timing error, detuning error, and decoherence. We hope the protocol may provide a method for preparing stable entangled cat states.
\end{abstract}

\maketitle

\section{Introduction}\label{I}

Quantum computers, which utilize entanglement superposition properties for information prosessing, possess the potential to outperform classical computers on some certain problems \cite{MA2000,Steane1998,annurev:/content/journals/10.1146/annurev-conmatphys-031119-050605,Arute2019,PhysRevA.72.050306}, such as searching unsorted databases \cite{PhysRevA.72.050306}.
However, quantum computers also face many challenges. For example, the experimental operating and environment noise may cause errors, especially on large-scale quantum computing \cite{PhysRevA.86.032324,Litinski2019gameofsurfacecodes}. Thus, quantum computing and its scaling will be limited.
To reduce error rate, protocols for quantum error correction have been developed in the past decades \cite{annurev:/content/journals/10.1146/annurev-conmatphys-031119-050605,PhysRevA.52.R2493,KITAEV20032,PRXQuantum.3.010329,Devitt_2013,PhysRevA.97.022335,PhysRevX.9.041009}. Encoding quantum information onto bosonic systems is beneficial to quantum error correction in some aspects \cite{CAI202150,PhysRevX.9.041053,MA20211789,PhysRevA.68.042319,PhysRevApplied.18.024076,Yang2023,PhysRevX.6.031006}.
For example, bosonic systems can provide infinitely large Hilbert space, which could be used to effectively protect and process quantum information \cite{CAI202150,PhysRevX.9.041053}.

Cat-state qubits are one kind of the promising bonsic quantum qubits in quantum information processing \cite{PhysRevA.68.042319,PhysRevA.104.013715,PhysRevLett.114.100403,ShiBiaoZheng1998,PhysRevLett.119.030502}, whose two logical qubits are usually represented by two orthogonal cat states (i.e., superposition of coherent states) \cite{AGilchrist_2004,PhysRevLett.126.023602,PhysRevResearch.4.013233,PhysRevA.106.042430}. Cat-state qubits have some unique advantages. For instance, they are noise biased \cite{MA20211789,CAI202150}. The phase-flip error can be effectively suppressed, only the bit-flip error needs to be concerned in error correction. Thus, the number of building blocks for error correction can be significantly reduced \cite{PhysRevX.9.041053,Mirrahimi_2014,PhysRevX.9.041009,PhysRevApplied.18.024076}. In addition, cat qubits have an enhanced lifetime with error corrections \cite{Ofek2016,PhysRevA.109.022437}. {Therefore, cat-state qubits have received much attention and many protocols \cite{PhysRevResearch.4.013233,PhysRevA.109.022437,PhysRevResearch.4.013233,Chen2024Prl,PhysRevLett.114.193602,Zhang:21,lpor.202200336,Ourjoumtsev2007,lpor.202300103} in view of cat-state qubits have been proposed, such as experimental preparation squeezed cat states \cite{lpor.202200336}, one-step parity measurement of $N$ cat-state qubits \cite{PhysRevA.109.022437} and the remote preparation of cat state \cite{lpor.202300103}.} {Entangled cat states are also important for the demonstration of the fundamentals of quantum physics, and have wide applications in modern quantum technologies \cite{PhysRevApplied.18.024076,Yang2023,Zhang2020,PhysRevLett.126.023602,PhysRevLett.126.023602,Yang:18}. } Protocols have been presented for producing entangled cat states, such as the generation of giant entangled cat states \cite{PhysRevLett.126.023602} and the preparation of multidimensional entangled cat states in cavity quantum electrodynamics (QED) \cite{Yang2023,Zhang2020}.

Among the various methods for preparing entangled cat states, the M${\o}$lmer-S${\o}$rensen (MS) entangling gate is a commonly employed technique \cite{PhysRevLett.126.023602,PhysRevApplied.18.024076}. The MS entangling gate has some specific characteristics \cite{PhysRevLett.82.1971,PhysRevA.62.022311}. Firstly, it makes qubits independent of the decoherence of the motion mode  \cite{PhysRevLett.82.1971,PhysRevA.62.022311,PhysRevA.101.030301}. Secondly, it possesses a built-in noise-resilience feature against certain types of local noise \cite{PhysRevApplied.18.024076}.
Therefore, the MS entangling gate inspired a wide range of research interests \cite{Shi-LiangZhu2006,PhysRevLett.112.190502,PhysRevLett.114.120502,PhysRevLett.117.060504,PhysRevLett.117.060505,PhysRevLett.119.220505,PhysRevA.101.030301,PhysRevA.72.050306}.
For example, in 2021, Chen $et$ $al$. used cat-code MS gate to prepare entangled cat states \cite{PhysRevApplied.18.024076}. But the protocol \cite{PhysRevApplied.18.024076} could be sensitive to timing error. We note that some kinds of composite pulses can play a role in robust control of quantum system and reduce errors \cite{Haddadfarshi_2016,PhysRevLett.121.180501,PhysRevLett.121.180502,WANG2021127033,PhysRevA.95.022328,PhysRevLett.119.220505,PhysRevLett.120.020501}. For example, Haddadfarshi $et$ $al$. pointed out that properly designed polychromatic control pulses can greatly reduce errors \cite{Haddadfarshi_2016} and then Webb $et$ $al$. demonstrated the technique experimentally \cite{PhysRevLett.121.180501}. Shapira $et$ $al$. generalized the MS entangling gate by using additional frequency components in the laser drive \cite{PhysRevLett.121.180502}. Wang $et$ $al$. used suitably designed laser pulse with modulated amplitude and phase to improve the fidelity of the MS entangling gate \cite{WANG2021127033}.

Inspired by the literatures above in this paper, we propose to use cat-code MS gate and composite pulses to prepare high-fidelity entangled cat states. The composite pulses are constructed by two-photon drives with different amplitudes and frequencies. Under such a drive, we can obtain more stable entangled cat states with appropriate parameters. Meanwhile, the sensitivities of the entangled cat states against errors are significantly reduced, and the robustness to decoherence are obviously improved.

The article is organized as follows. In Sec.~\ref{II}, we describe the physical model, and give the concrete form of the Hamiltonian. The effective Hamiltonian is obtained by setting appropriate parameters, and the maximum entangled cat states are prepared with the MS entangling gate. In Sec.~\ref{III}, we introduce the composite drives. Then we prepare the entangled cat states under the effective Hamiltonian, and numerically simulate the feasibility of the protocol. In Sec.~\ref{IV}, we evaluate the performance of the protocol against errors, and discuss the fidelity of the entangled cat states in the presence of the dephasing and single-photon loss. Finally, conclusions are given in Sec.~\ref{V}.

\section{Cat-code MS entangling gate} \label{II}
The physical model consists of two Kerr-nonlinear cavities ($\mathcal{A}_1,\mathcal{A}_2$) with the same frequency $\omega_k$, and another cavity ($\mathcal{A}_0$) with frequency $\omega_0$. The Hamiltonian of this system in the interaction picture is given by (assume $\hbar=1$)
\begin{eqnarray}\label{1}
H=\sum_{k=1,2}H_k^{\mathrm{Kerr}}+H_{I}.
\end{eqnarray}
The interaction Hamiltonian $H_{I}$ is \cite{PhysRevApplied.18.024076}
\begin{eqnarray}\label{2}
H_I=\sum_{k=1,2}Ja_{k}a^{\dag}_{0}e^{i\Delta t}+\mathrm{H.c.},
\end{eqnarray}
where $J$ is the coupling strength, $a_k(k=1,2)$ is the annihilation operator of the Kerr-nonlinear resonator, $a_0$ is the annihilation operator of the cavity, and $\Delta=\omega_0-\omega_k$ is the detuning. Each Kerr-nonlinear resonator is resonantly driven by a single-mode two-photon squeezing drive with frequency $\omega_p=2\omega_k$ and amplitude $\Omega_p$. The Hamiltonian of the Kerr parametric oscillators (KPOs) $H_k^{\mathrm{Kerr}}$ in Eq.~(\ref{1}) is \cite{PhysRevX.9.041009,Puri2017}
\begin{eqnarray}\label{3}
H_k^{\mathrm{Kerr}}=-Ka^{\dag2}_k a^{2}_k+(\Omega_p a^{2}_k+\Omega^{*}_p a^{\dag2}_k),
\end{eqnarray}
where $K$ is Kerr nonlinearity coefficient \cite{PhysRevA.93.013808,PhysRevA.86.013814}. The Hamiltonian $H_k^{\mathrm{Kerr}}$ can be rewritten as:
\begin{eqnarray}\label{4}
H_k^{\mathrm{Kerr}}=-K(a^{\dag 2}_k-\frac{\Omega^*_p}{K})(a_k^2-\frac{\Omega_p}{K})+\frac{|\Omega_p|^2}{K}.
\end{eqnarray}
Equation~(\ref{4}) clearly shows that when $\alpha=\sqrt{\frac{\Omega_p}{K}}$, the two coherent states $|\pm\alpha\rangle$ are degenerate eigenstates of the annihilation operator $a_k$. Therefore, the superpositions of the coherent states \cite{CAI202150,PhysRevX.9.041009,Puri2017,Puri2020,PhysRevX.14.031040}
\begin{eqnarray}\label{5}
|C_{\pm}\rangle_k=N_{\pm}(|\alpha\rangle_k\pm |-\alpha\rangle_k),
\end{eqnarray}
are also the degenerate eigenstates of $H_k^{\mathrm{Kerr}}$. Here $N_{\pm}$ are normalization coefficients. For simplicity, we set the parameters $\{K,\Omega_p,J,\Delta\}>0$ and $\alpha=\alpha^{*}>0$.
\begin{figure}
  \centering
  \subfigure{{\includegraphics[width=0.8\linewidth]{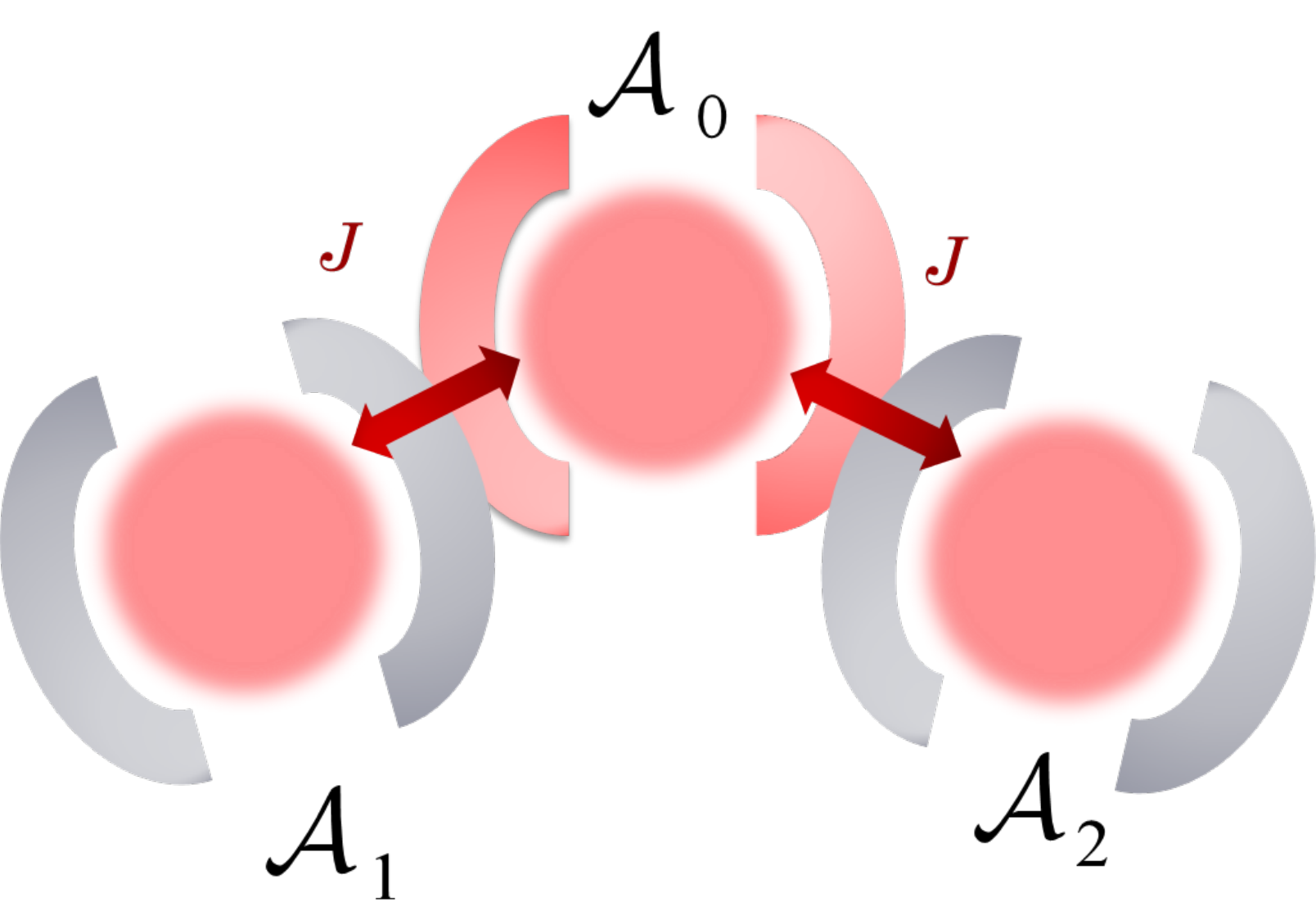}}\label{1a}}
  \caption{Two Kerr-nonlinear resonators ($\mathcal{A}_1$, $\mathcal{A}_2$) are coupled to a cavity ($\mathcal{A}_0$) with a coupling strength of $J$, and each Kerr nonlinear resonator is driven by a two-photon drive with a frequency $\Omega_p$.}
\label{F1}
\end{figure}

The subspace composed by $|C_{\pm}\rangle_k$ is called the cat-state subspace, which is separated from the other subspaces with an energy gap $E_{\mathrm{gap}}\simeq4K\alpha^2$ \cite{PhysRevApplied.18.024076}.
For large $\alpha$, the annihilation operator $a_k$ only causes the flip between the two cat states, i.e.,
\begin{eqnarray}\label{6}
a_k|C_{\pm}\rangle_k\simeq\alpha|C_{\mp}\rangle_k.
\end{eqnarray}
When the condition $E_{\mathrm{gap}}\gg J$ is satisfied, the transition probability from the ground states $|C_{\pm}\rangle_k$ to the excited states is extremely small \cite{PhysRevApplied.18.024076}.
Therefore, the dynamics of the system can be restricted in the cat-state subspace with the effective Hamiltonian
\begin{eqnarray}\label{7}
H_{\mathrm{eff}}&\approx&\sum_{k=1,2}\frac{\Omega^{2}_{p}}{K}(|C_{-}\rangle_{k}\langle{C_{-}}|+|C_{+}\rangle_{k}\langle{C}_{+}|)\cr
&&\cr
&&+J\alpha[|C_{+}\rangle_{k}\langle{C}_{-}|(a_{0}e^{-i\Delta{t}}+a^{\dag}_{0}e^{i\Delta{t}})+\mathrm{H.c.}].
\end{eqnarray}
The first term in the right-hand side of Eq.~(\ref{7}) can be dropped because it is the identity matrix of the cat-state subspace. Defining Pauli matrices $\sigma^{+}_k=|C_-\rangle_k\langle C_+|$ and $\sigma^{-}_k=|C_+\rangle_k\langle C_-|$, the Hamiltonian in Eq.~(\ref{7}) reduces to \cite{PhysRevApplied.18.024076,Grimm2020,arXiv:2407.10940}
\begin{eqnarray}\label{8}
H_{\mathrm{eff}}=2J\alpha S_x(a_0e^{-i\Delta t}+a^{\dag}_0 e^{i\Delta t}),
\end{eqnarray}
where $S_x=\frac{1}{2}\sum_{k=1,2}(\sigma^{+}_k+\sigma^{-}_k)$.

To well understand the evolution of the system, the evolution operator based on the effective Hamiltonian $H_{\mathrm{eff}}$ can be calculated with the Magnus expansion ({see Appendix \ref{AA} for details}) \cite{Magnus1954OnTE,Blanes_2010,BLANES2009151}:
\begin{eqnarray}\label{9}
U(t)=\exp\{-i[(\chi(t)a^{\dag}_{0}+\mathrm{H.c.})S_{x}+\beta(t)S^{2}_{x}]\},
\end{eqnarray}
\begin{eqnarray}\label{10}
\chi(t)&=&\frac{2iJ\alpha}{\Delta}[1-\exp(i\Delta{t})],\cr
&&\cr
\beta(t)&=&(\frac{2J\alpha}{\Delta})^{2}(\sin\Delta{t}-\Delta{t}).
\end{eqnarray}
Note that a large detuning $\Delta$ is necessary for the Magnus expansion to proceed. Setting $t=T=2\pi/\Delta$, there have $\chi(T)=0$ and $\beta(T)=(\frac{2J\alpha}{\Delta})^{2}\cdot(-2\pi)$. When choosing the parameter $\Delta=4J\alpha$, then $\beta(T)=-\frac{\pi}{2}$. Thus the evolution operator at time $T$ reads
\begin{eqnarray}\label{11}
U(T)=\exp(i\frac{\pi}{2}S_{x}^{2}).
\end{eqnarray}
The $U(T)$  in Eq.~(\ref{11}) is the well-known MS entangling gate \cite{PhysRevA.62.022311}. {The MS entangling gate is an important physical resource in quantum computing, for example, it can be applied to prepare entangled states
\cite{PhysRevApplied.18.024076,PhysRevLett.82.1971}.} Under the action of $U(T)$, we can realize the preparation of the entangled cat states. As shown in Eq.~(\ref{12}), the left side of Eq.~(\ref{12}) are the four different initial states (the input states), and the right side are the corresponding maximum entangled cat states (the output states) obtained via $U(T)$ \cite{PhysRevApplied.18.024076,PhysRevLett.82.1971}.
\begin{eqnarray}\label{12}
|C_+\rangle|C_+\rangle|0\rangle&&\rightarrow\!\frac{1}{\sqrt{2}}(|C_+\rangle|C_+\rangle|0\rangle+i|C_-\rangle|C_-\rangle|0\rangle),\cr\cr
|C_-\rangle|C_-\rangle|0\rangle&&\rightarrow\!\frac{1}{\sqrt{2}}(|C_-\rangle|C_-\rangle|0\rangle+i|C_+\rangle|C_+\rangle|0\rangle),\cr\cr
|C_+\rangle|C_-\rangle|0\rangle&&\rightarrow\!\frac{1}{\sqrt{2}}(|C_+\rangle|C_-\rangle|0\rangle-i|C_-\rangle|C_+\rangle|0\rangle),\cr\cr
|C_-\rangle|C_+\rangle|0\rangle&&\rightarrow\!\frac{1}{\sqrt{2}}(|C_-\rangle|C_+\rangle|0\rangle-i|C_+\rangle|C_-\rangle|0\rangle).
\end{eqnarray}
For convenience, the initial state of the system is set as $|\Psi_{in}\rangle=|C_{+}\rangle|C_{+}\rangle|0\rangle$, and the corresponding entangled cat state is $|\Psi_{out}\rangle=\frac{1}{\sqrt{2}}(|C_+\rangle|C_+\rangle|0\rangle+i|C_-\rangle|C_-\rangle|0\rangle)$ in this manuscript. All numerical simulation results in this section are obtained by using the full Hamiltonian given by Eq.~(\ref{1}). Here, the fidelity of entangled state is given by $F=\langle\Psi|\rho|\Psi\rangle$, where $|\Psi\rangle$ is the target entangled cat state i.e., $|\Psi_{out}\rangle$, and $\rho$ is the density operator. The experimental parameters are chosen as $\alpha=2$, $K/2\pi=20\mathrm{MHz}$, $\Omega_{p}=K\alpha^2$, $J=2\pi\mathrm{MHz}$, and $\Delta=4J\alpha$ \cite{PhysRevApplied.18.024076}.
\begin{figure}
\centering
\subfigure{\includegraphics[width=9cm]{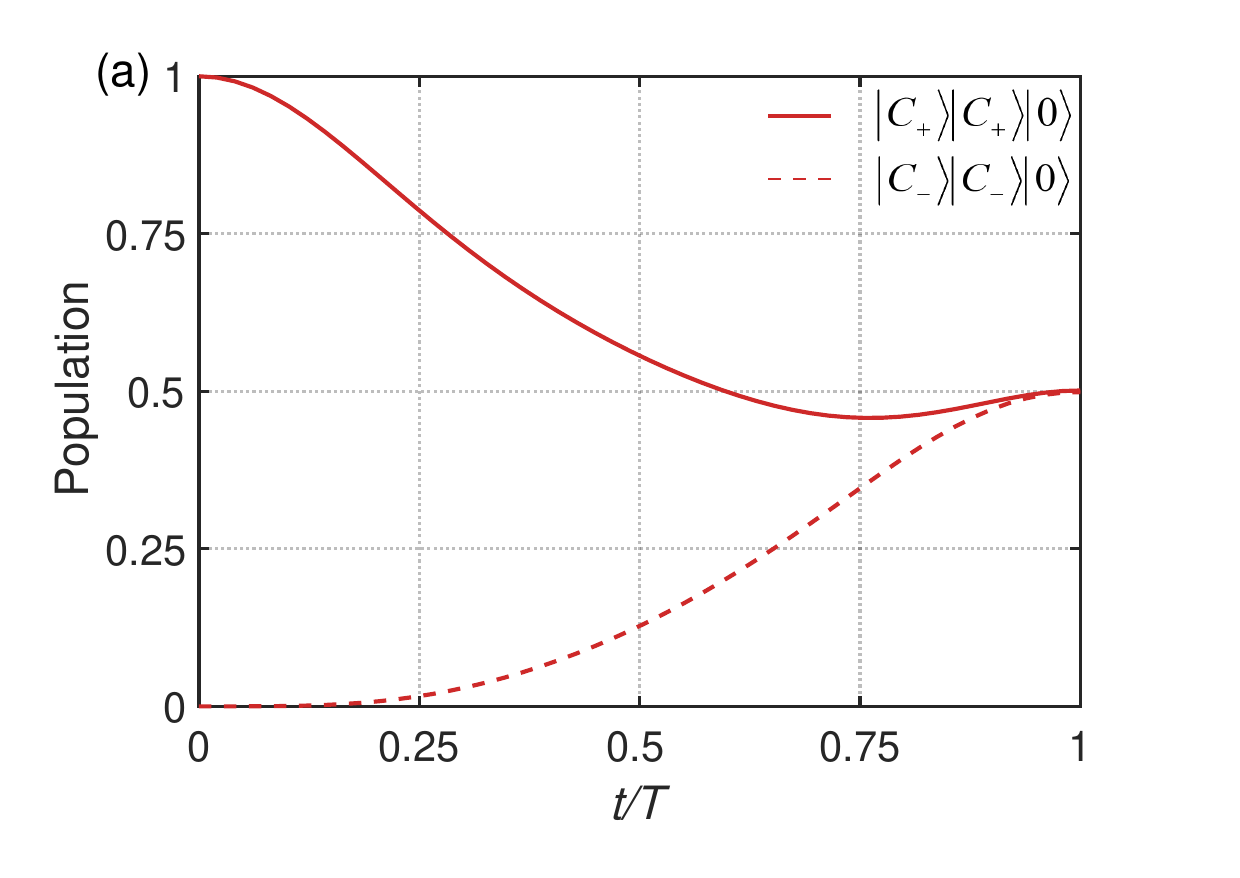}\label{2a}}
\subfigure{\includegraphics[width=9cm]{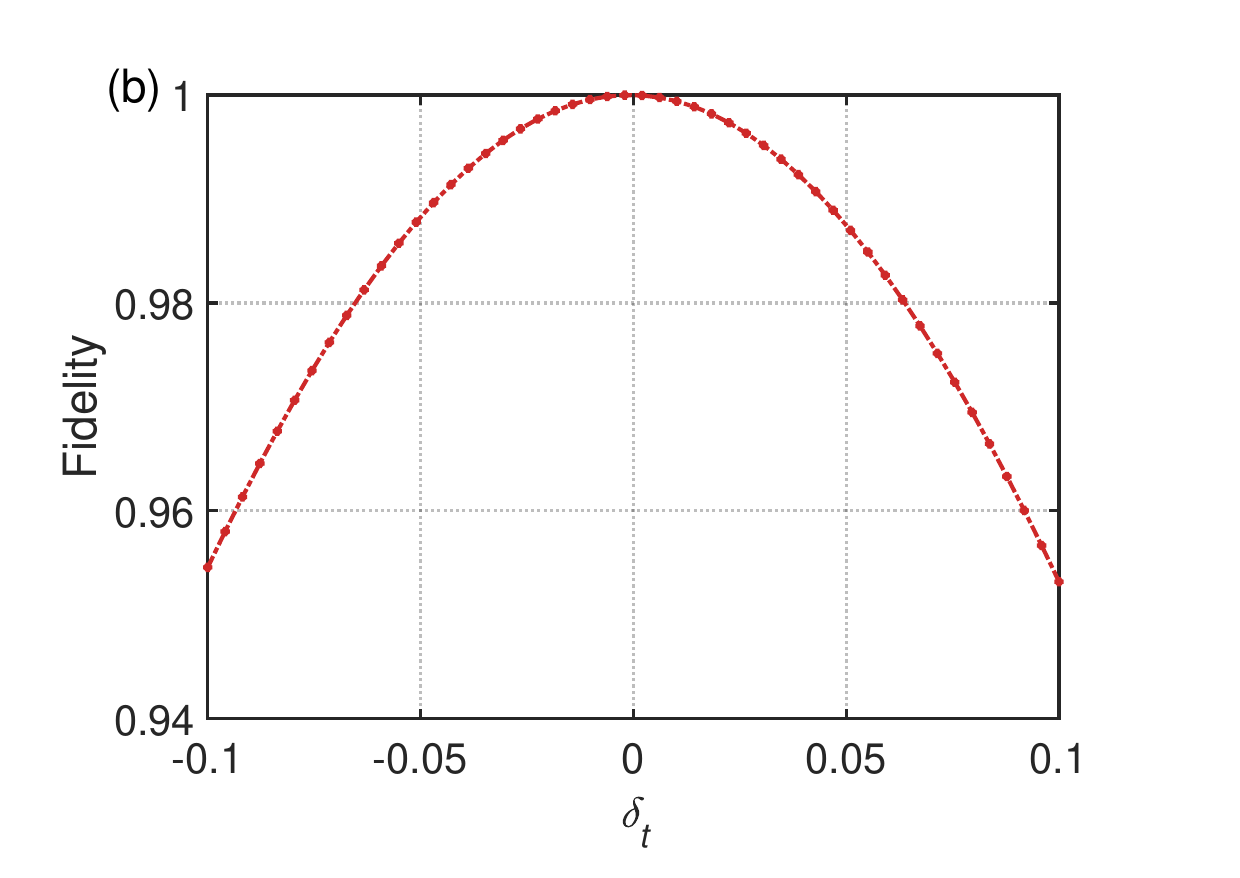}\label{2b}}
\caption{(a) Population envolution of the cat states. The solid line represents the state $|C_+\rangle|C_+\rangle|0\rangle$, and the dotted line represents the state $|C_-\rangle|C_-\rangle|0\rangle$. (b) The fidelity against timing error $\delta_t$, error range is selected for [-0.1, 0.1]. The experimental parameters are set as $\alpha=2$, $T=2\pi/\Delta$, the Kerr nolinearity $K/2\pi=20\mathrm{MHz}$, the coupling strength $J/2\pi=1\mathrm{MHz}$ the driving amplitude $\Omega_{p}=K\alpha^2$, and the detuning $\Delta$ should obey $\Delta=4J\alpha$ \cite{PhysRevApplied.18.024076}.}\label{F2}
\end{figure}

{In Fig.~\ref{2a}, we plot the population evolutions with time of the states $|C_{+}\rangle|C_{+}\rangle|0\rangle$ and $|C_{-}\rangle|C_{-}\rangle|0\rangle$. As it is shown, the populations of both states are close to $50\%$ around the gate time $T$.
That is, the maximum entangled cat state can be successfully prepared when $t=T$. However, the maximum entangled cat state can only exist when time is extremely close to $T$. To clearly see the fidelity of the maximum entangled cat state against timing errors, we plot Fig.~\ref{2b}.} When the timing error $\delta_t$ deviates from 0, i.e., the actual operation time deviates from $T$, the fidelity of the entangled cat state $|\Psi_{out}\rangle$ drops significantly. Here, the relative error of the parameter $x$ is defined via $\delta_x=(x'-x)/x$, in which $x'$ denotes the actual value and $x$ is the ideal value.

\section{Preparation of high-fidelity entangled cat state} \label{III}

As shown in the second section, the fidelity of the entangled cat state is relatively sensitive to time error. In order to obtain stable and high-fidelity entangled cat state, we introduce composite two-photon squeezing drives with frequency $\omega_n$ and amplitude $\Omega_n$. Then the total Hamiltonian $H$ in Eq.~(\ref{1}) changes to
\begin{eqnarray}\label{13}
H'=\sum_{k=1,2}H_k^{\mathrm{Kerr}'}+H_I,
\end{eqnarray}
\begin{eqnarray}\label{14}
H_k^{\mathrm{Kerr}'}=&-&Ka^{\dag2}_k a^{2}_k+(\Omega_p a^{2}_k+\Omega^{*}_pa^{\dag2}_k)\cr
&&\cr
&+&\sum_{n=1}^N(\Omega_n a^{\dag2}_k e^{i\delta_n t}+\Omega_n^*a^2_k e^{-i\delta_n t}).
\end{eqnarray}
{The last term in Eq.~(\ref{14}) is the additional composite drives, where $\delta_n=\omega_n-\omega_k$ is the detuning of the $n$th two-photon drive, and $N$ represents the number of two-photon drives that we may introduce for the stable entangled cat states.} According to effective Hamiltonian theory \cite{doi:10.1139/p07-060}, the effective Hamiltonian in the limit of large detunings can be derived as
\begin{eqnarray}\label{15}
H_{\mathrm{eff},N}(t)=\sum_{n=1}^N\frac{4J\Omega_n}{\varepsilon_n}
[a_k^{\dag}a_0^{\dag}e^{i(\Delta-\delta_n)t}+\mathrm{H.c.}],
\end{eqnarray}
where $\frac{1}{\varepsilon_n}=({\frac{1}{\Delta}+\frac{1}{\delta_n}})$.
Similar to the derivation from Eq.~(\ref{1}) to Eq.~(\ref{8}), we can get a new effective Hamiltonian by restricting the Hamiltonian $H_{\mathrm{eff},N}(t)$ to the cat-state subspace as
\begin{eqnarray}\label{16}
H_{\mathrm{eff},N}'(t)&=&2J\alpha S_x\sum_{n=1}^N\frac{4\Omega_n}{\varepsilon_n}
[a_0^{\dag}e^{i(\Delta-\delta_n)t}+\mathrm{H.c.}].
\end{eqnarray}
For simplicity, we define $\frac{4\Omega_n}{\varepsilon_n}=r_n$ and $\zeta=(\Delta-\delta_n)/n$. Then the effective Hamiltonian $H_{\mathrm{eff},N}'(t)$ is expressed as
\begin{eqnarray}\label{17}
H_{\mathrm{MS},N}(t)=2J\alpha S_x\sum_{n=1}^N r_n(a_0e^{in\zeta t}+a_0^{\dag} e^{-in\zeta t}).
\end{eqnarray}
\begin{figure}
  \centering
  \includegraphics[width=1.1\linewidth]{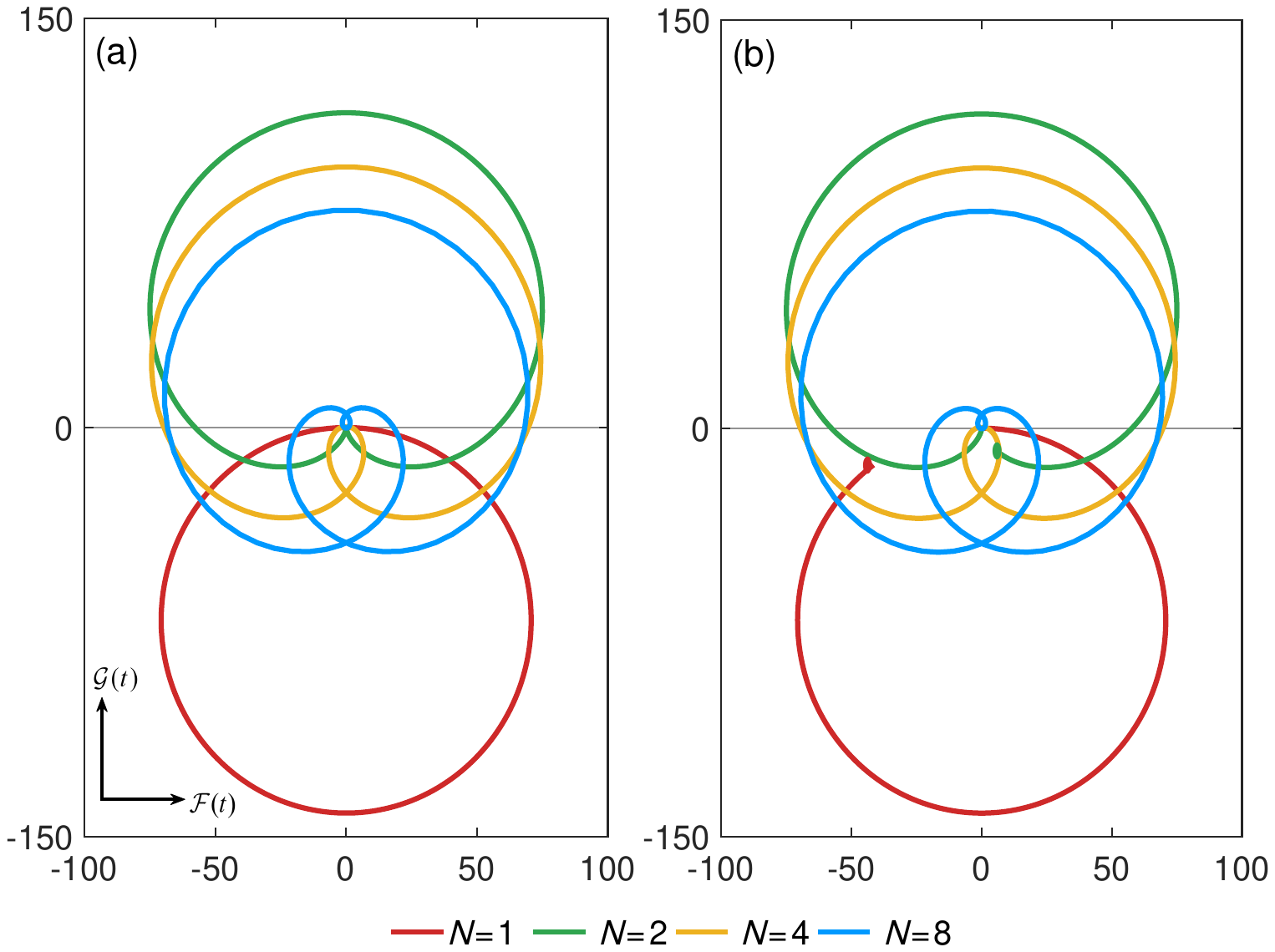}
  \caption{Phase-space trajectories of $\mathcal{F}(t)$ and $\mathcal{G}(t)$ for $N\in[1,2,4,8]$. (a) without errors, (b) with $10\%$ timing error.}\label{F3}
\end{figure}

To understand the construction of composite pulses, we introduce the dimensionless position operator $x=\frac{1}{\sqrt{2}}(a_0+a_0^{\dag})$  and the momentum operator $p=\frac{i}{\sqrt{2}}(a_0^{\dag}-a_0)$. The Hamiltonian in Eq.~(\ref{17}) is reformulated as
\begin{eqnarray}\label{18}
\mathcal{H}_{\mathrm{MS},N}(t)=f(t)S_x x+g(t)S_x p,
\end{eqnarray}
where $f(t)=2\sqrt{2}J\alpha\sum_{n=1}^N\cos(n\zeta t)$ and $g(t)=-2\sqrt{2}J\alpha\sum_{n=1}^N\sin(n\zeta t)$. The propagator for the Hamiltonian in Eq.~(\ref{18}) can be expressed as \cite{PhysRevLett.121.180502,PhysRevA.62.022311}
\begin{eqnarray}\label{19}
\mathcal{U}(t)=e^{-i\mathcal{F}(t)S_x x}e^{-i\mathcal{G}(t)S_x p}e^{-i\mathcal{A}(t)S_x^2},
\end{eqnarray}
where
\begin{eqnarray}\label{20}
\mathcal{F}(t)=&&\int_0^t\!f(t')dt'=\frac{2\sqrt{2}J\alpha}{\zeta}\sum_{n=1}^N\!\frac{r_n}{n}\sin(n\zeta t),\cr\cr
\mathcal{G}(t)=&&\int_0^t\!g(t')dt'=\frac{2\sqrt{2}J\alpha}{\zeta}\sum_{n=1}^N\!\frac{r_n}{n}[\cos(n\zeta t)-1],\cr\cr
\mathcal{A}(t)=&&-\int_0^t\!\mathcal{F}(t')g(t')dt',
\end{eqnarray}
and $\{r_n\}^N_{n=1}$ is a numerical set of $r_n$ when $N$ takes different values \cite{PhysRevLett.121.180502}
\begin{eqnarray}\label{21}
r\!_n\!=\!(-1)^{N-n}\frac{N!}{2^N}\!\sqrt{\frac{2\sqrt{\pi}}{(\!N\!-\!1\!)\Gamma(\!N\!+\!\frac{1}{2}\!)}}
\!\frac{(N-1)!}{(\!N\!-\!n\!)!(\!n\!-\!1\!)!}.
\end{eqnarray}
Here $\Gamma(a)=\int_0^\infty e^{-x}x^{a-1}dx$.

As seen in Eq.~(\ref{19}), the evolution of the operator $\mathcal{U}(t)$ is related to the trajectorys with dimensionless coordinates $\mathcal{F}(t)$ and $\mathcal{G}(t)$ in the $x-p$ phase-space. Equation~(\ref{20}) shows that $\mathcal{F}(t)$ and $\mathcal{G}(t)$ are respectively the integrations of $f(t)$ and $g(t)$ \cite{PhysRevA.62.022311,Haddadfarshi_2016}. Ideally, when the time $t=\tau=\frac{2\pi}{\zeta}$, $\mathcal{F}(\tau)=0$, $\mathcal{G}(\tau)=0$, and the phase-space trajectorys of $(\mathcal{F}(t), \mathcal{G}(t))$ is closed as seen in Fig.~\ref{F3}(a). Meanwhile, the operator  $\mathcal{U}(\tau)=e^{-i\mathcal{A}(\tau)S_x^2}$, that is, only the internal spin evolution is retained. However, errors will lead to the phase-space trajectory not closed completely, which in turn may causes infidelity of entangled cat states \cite{PhysRevLett.83.5166,WANG2021127033,PhysRevLett.121.180501}. As we can see from Fig.~\ref{F3}(b), the circular trajectory becomes more complete with the increase of $N$. In other words, we can reduce errors and improve the fidelity of the scheme by increasing $N$.

\begin{figure}
\includegraphics[width=1.2\linewidth]{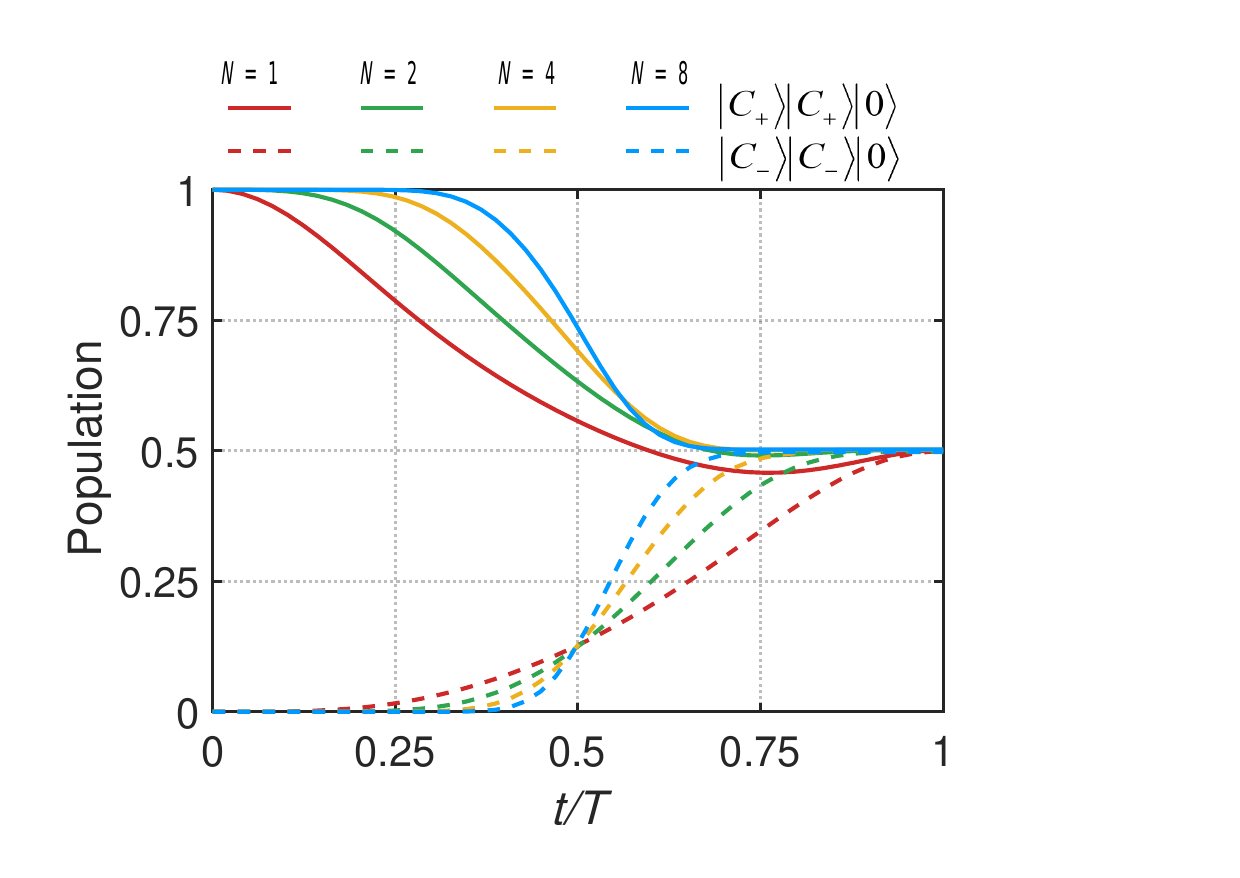}
\caption{The population evolution of cat states, the values of $N\in1,2,4,8$. The solid line represents the state $|C_+\rangle|C_+\rangle|0\rangle$, and the dotted line represents the state $|C_-\rangle|C_-\rangle|0\rangle$.}\label{F4}
\end{figure}
The full Hamiltonian of the optimized protocol is written as
\begin{eqnarray}\label{22}
H_N(t)=&&\sum_{k=1,2}[-Ka^{\dag2}_k a^{2}_k+
(\Omega_p a^{2}_k+\Omega^{*}_p a^{\dag2}_k)]\cr
&&\cr
&&+\sum_{k=1,2}[J_N(t)a_{k}a^{\dag}_{0}\exp(i\Delta t)+\mathrm{H.c.}],
\end{eqnarray}
where $J_N(t)=J e^{-i\Delta t}\sum_{n=1}^{N}r_n e^{in\zeta t}$ is the composite drive. If the condition $\Delta=\zeta$ is satisfied, $J_1(t)=J$, the case $N=1$ corresponds to the unoptimized protocol, which is disscussed in the second section.

{In Fig.~\ref{F4}, we plot the populations of the cat states with different $N$ ($N=1, 2, 4, 8$). The curves with $N=1$ correspond to the curves in Fig. 2(a), i.e., the populations of the cat states under the cat-code MS gate \cite{PhysRevApplied.18.024076}. As shown in Fig.~\ref{F4}, compared to the results with $N=1$, the entangled cat states can exist for a longer duration when $N=2, 4, 8$. This means that more stable entangled cat states can be obtained by simply increasing $N$.} Here, the parameters $\{\alpha, K, \Omega_p, \Delta\}$ are same as in Fig.~\ref{F2}, and $\zeta=\Delta$.
\begin{figure}
\includegraphics[width=1\linewidth]{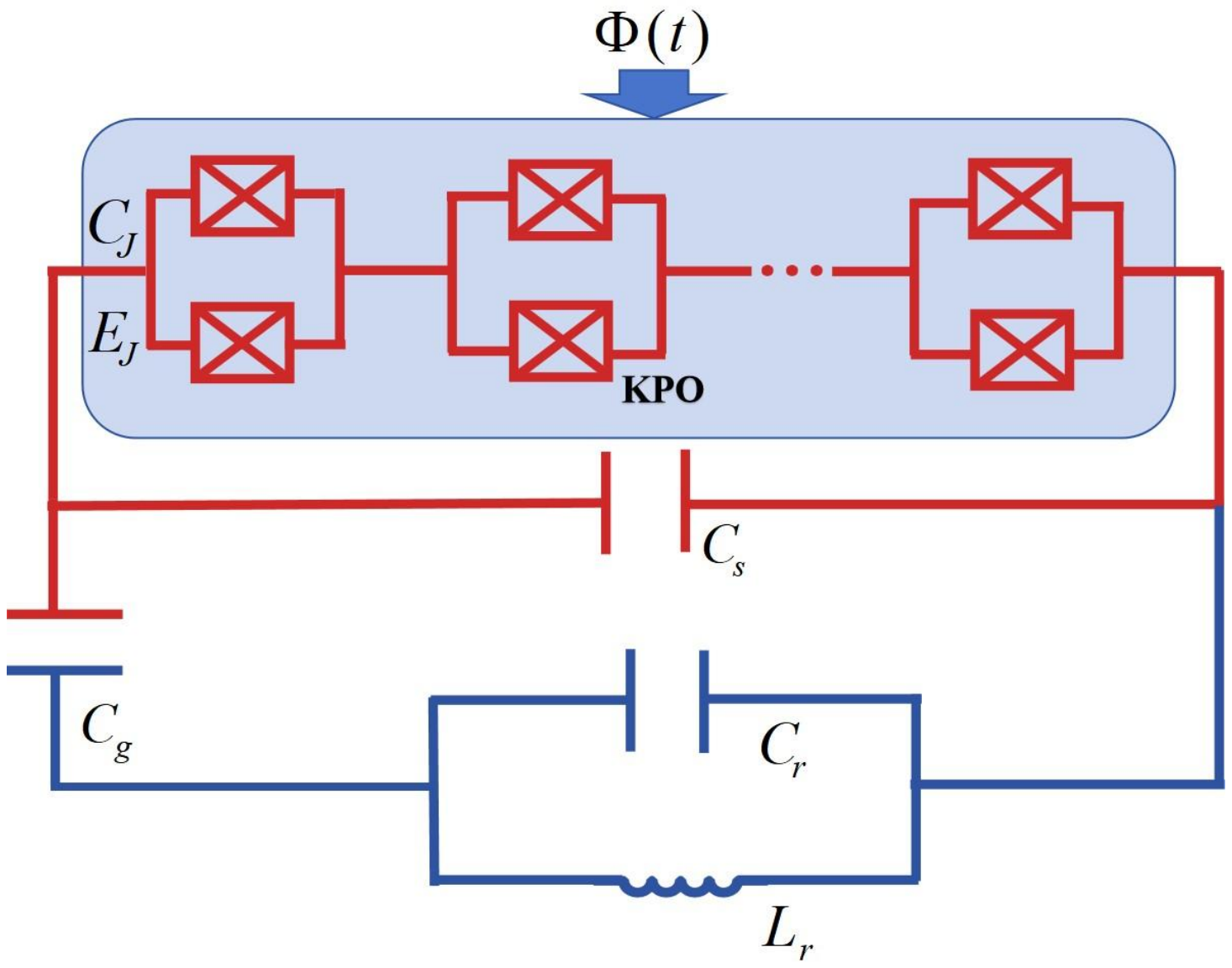}
\caption{The effective circuit diagram of the protocol. The shaded area which is an array of Josephson junctions represents the KPO, and the $LC$ oscillator represents the cavity mode $\mathcal{A}_0$ in our protocol. The KPO coupled to the $LC$ oscillator via capacitor $C_g$.  Here, $C_J$ and $E_J$ correspond to the capacitance and energy of the Josephson junctions array, respectively, and $\Phi(t)$ represents the external magnetic flux. $C_s$ and $C_g$ are two additional large capacitors.}\label{F5}
\end{figure}

{Superconducting quantum interference devices (SQUIDs) can be a possible implementation of the protocol \cite{PhysRevA.67.042311,PhysRevB.87.144301,GU20171,PhysRevX.9.021049,PhysRevB.75.140515,PhysRevLett.114.090503,PhysRevB.87.184501}. } As shown in Fig.~\ref{F5}, considering the effective circuit of the array of Josephson junctions coupled to an $LC$ oscillator through a large gate capacitance $C_g$.
The array of Josephson junctions with the energy $E_J$ and capacitance $C_J$ serves as the KPO, \cite{PhysRevB.74.224506,PhysRevA.81.042304,PhysRevX.8.031007,PhysRevApplied.12.064037,RevModPhys.85.623}, and the $LC$ oscillator serves as the cavity mode $\mathcal{A}_0$ with the frequency $\omega_0=1/\sqrt{L_rC_r}$ \cite{PhysRevX.9.021049,Appl.2773988}.
The Josephson junction is shunted by an external large capacitance $C_s$.
Meanwhile, the Josephson energy $E_J$ can be adjusted by an external magnetic flux $\Phi(t)$ with frequency $\omega_p$, leading to $\tilde{E_J}=E_J+\delta E_J \cos(\omega_p t)$.

Utilizing the charge of the circuit and the external magnetic flux, one can construct the initial Hamiltonian that describes the circuit \cite{PhysRevA.76.042319}.
Then according to the standard quantization procedure of the circuit \cite{PhysRevB.75.140515,PhysRevLett.114.090503} and setting appropriate parameters, the Hamiltonian of a single KPO coupled to cavity mode $\mathcal{A}_0$ can be derived as \cite{PhysRevApplied.18.024076,PhysRevResearch.4.013233}
 \begin{eqnarray}\label{23}
\mathbf{H}=&&-Ka^{\dag2}a^2+\Omega_p(a^{\dag 2}+a^2)\cr\cr
&&+[J a a_0^{\dagger}e^{i\Delta t}+\mathrm{H.c.}].
\end{eqnarray}
The relevant parameters in Eq.~(\ref{23}) can be represented as follows \cite{PhysRevX.9.021049,PhysRevA.76.042319,PhysRevApplied.18.024076,PhysRevResearch.4.013233}:
$K=2E_C/K_0^2$, $\Omega_p=\delta E_J\omega_k/8E_J$, $J=-i2C_g eV_on_0/(C_g + C_s)$, and detuning $\Delta=\omega_0-\omega_k$. Where $E_C$ is the charging energy of the KPO, $K_0$ is the number of SQUIDs, $\omega_k=8\sqrt{E_cE_J/K_0}$ is the frequency of KPO, $V_o=\sqrt{\frac{\omega_0}{2C_r}}$ is the root-mean-square voltage of the local oscillator, $n_0= \sqrt[4]{E_J/(32K_0 E_C)}$ is the zero-point fluctuation. Considering the large distance between two KPOs, their coupling interaction can be neglected \cite{PhysRevX.9.021049,PhysRevA.76.042319,PhysRevApplied.18.024076,PhysRevResearch.4.013233}. Therefore, the Hamiltonian of two KPOs coupled to a cavity can be derived by summing up Eq.~(\ref{23}).

\begin{figure}
\centering
\subfigure{\includegraphics[width=9.5cm]{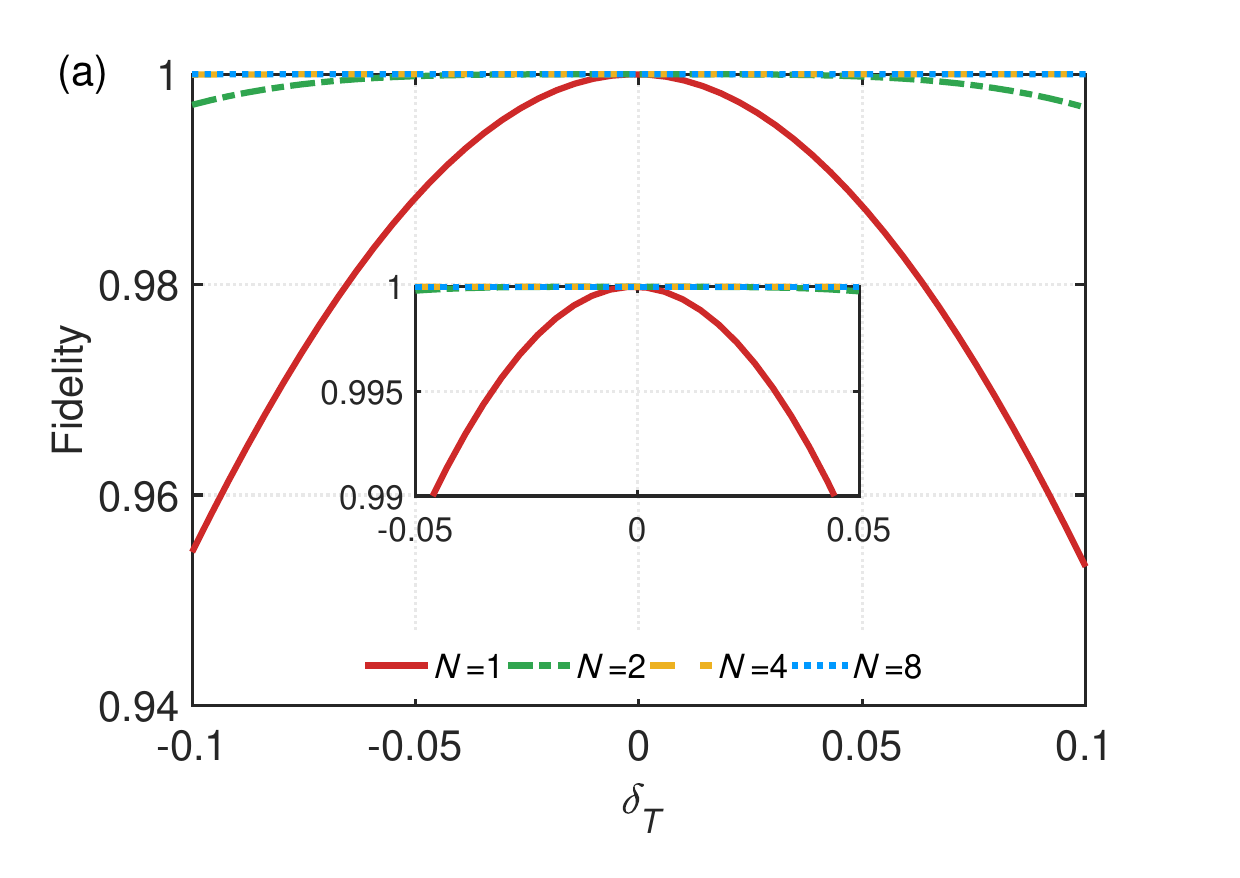}\label{T4}}
\subfigure{\includegraphics[width=9.5cm]{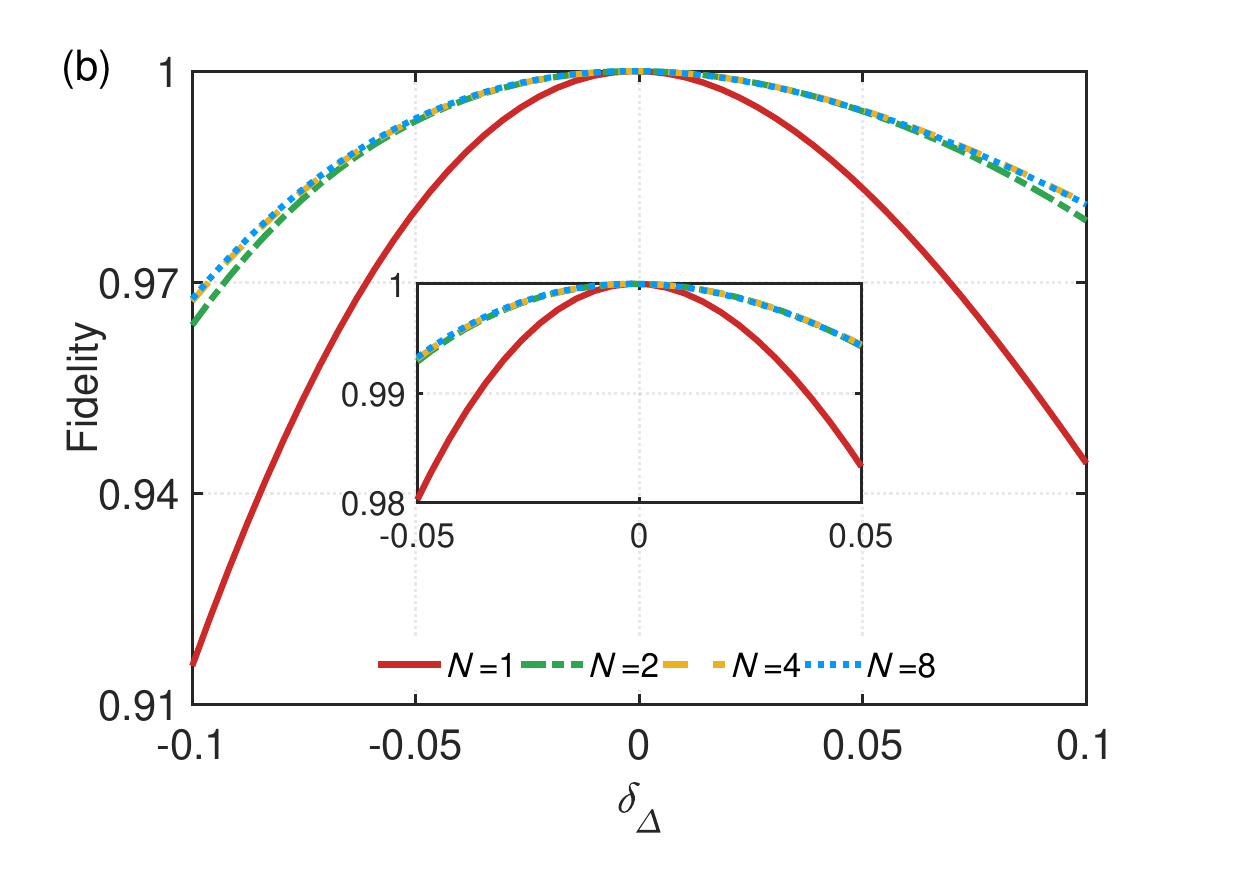}\label{D4}}
\subfigure{\includegraphics[width=9.5cm]{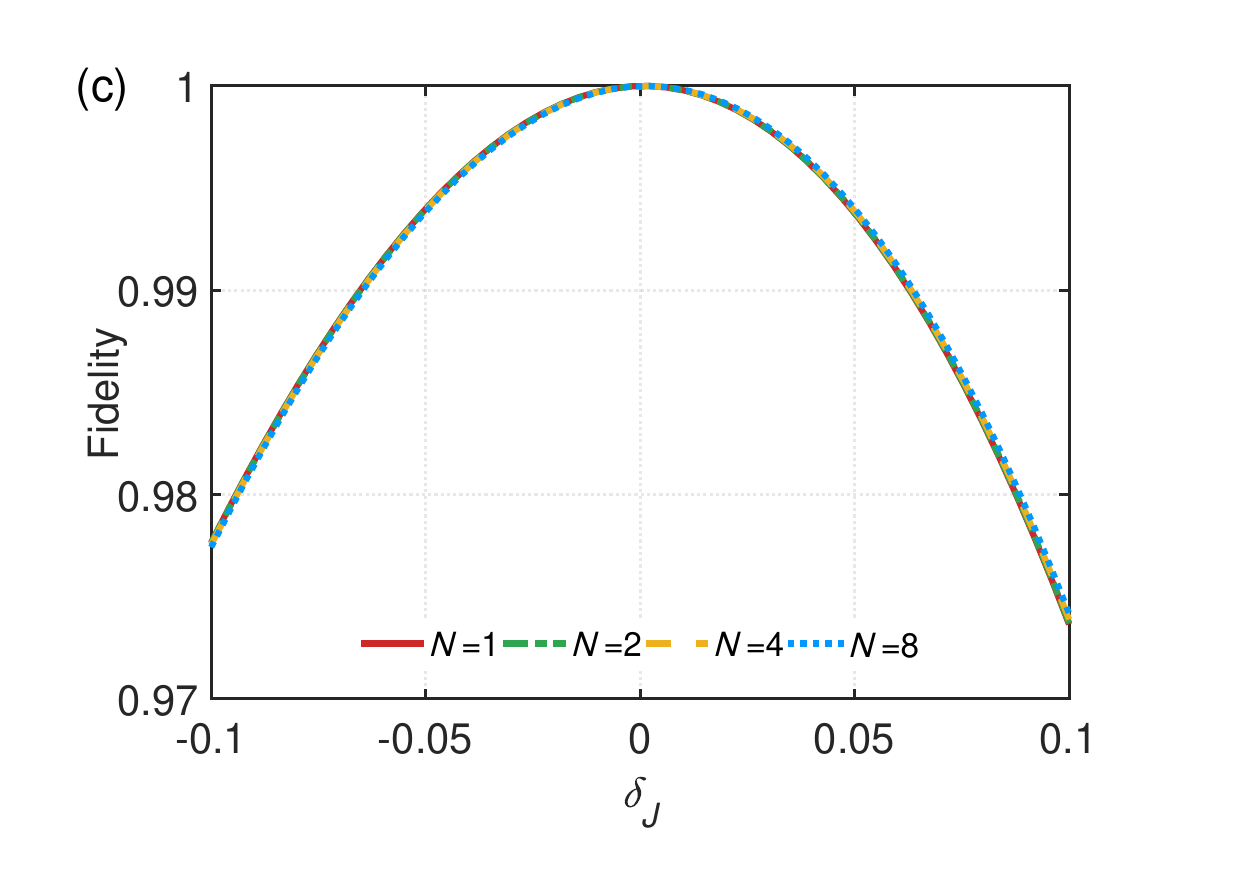}\label{J4}}
\caption{Fidelities vs parameter errors. (a) timing errors, (b) detuning errors, (c) coupling errors. In figures (a) and (b), increasing $N$ leads to a flatter response to control errors, the robustnesses against parameters errors are enhanced. The other parameters are the same as in Fig.~\ref{F4}.}
\end{figure}

\section{Error and decoherence analysis}\label{IV}
In this section, we will investigate the performance of the optimized protocol. Specifically, we discuss the effects of parameters errors on the fidelity of entangled cat state in the Sec.~\ref{A}. Then we analyze the robustness of the protocol to dephasing and single-photon losses in Sec.~\ref{B}. {Here, we choose the values $N\in[1,2,4,8]$.}
\subsection{Robustness against parametric errors}\label{A}
It is inevitable that imperfect experimental operations will cause infidelity in the entangled cat states. In this part, with the help of the numerical simulation, we will show the sensitivity of the protocol against parameter errors i.e., timing error $\delta_t$, detuning error $\delta_{\Delta}$, and coupling error $\delta_J$.

In Fig.~\ref{T4}, we numerically study the robustness of the  protocol for timing error $\delta_t$. The error range is $\delta_t\in[-0.1,0.1]$ and the curve for $N=1$ corresponds to the unoptimized one. { As it is shown, when $N=2$, the fidelities of the entangled cat states against timing errors have been greatly improved. Furthermore, when $N$ is set to larger values i.e., $N=4$ or $N=8$, the fidelities are further enhanced and are almost unaffected by timing errors. That is, with the increase of $N$, the robustnesses of the protocol to timing errors are enhanced.}

In actual experiments, imperfect experimental operations may cause inevitable detuning errors. We consider the range of detuning error $\delta_{\Delta}$ as $[-0.1, 0.1]$, and plot the fidelity versus detuning error $\delta_{\Delta}$ in the Fig.~\ref{D4}. {It can be seen from the figure that the fidelities of the entangled cat state for $N=2$ are much higher than that with $N=1$ when detuning errors exist. When $N$ increases, the fidelities become higher.} The inset in Fig.~\ref{D4} shows that when the error range is $[-0.05, 0.05]$, the fidelities are large than 0.99 when $N>1$ . This indicates that the increase of $N$ can reduce the sensitivity to detuning error.

Figure~\ref{J4} shows the robustness against coupling error $\delta_J$. {As shown in the figure, the variation of the coupling has a relatively small influence on the fidelities, and the fidelities against coupling errors are independent of the number of $N$. The reason can be seen from Eq.~(\ref{20}), where the coupling strength $J$ is separated from the summation term of $N$. Hence the fidelities of the entangled cat states against coupling errors are also independent of $N$.}

\subsection{Robustness against decoherence}\label{B}

In practical operations, decoherence is inevitable. For the system discussed, we consider two types of noise: single-photon losses and pure dephasing. The master equation of the system can be written as:
\begin{eqnarray}\label{24}
\dot{\rho}&=&-i[H,\rho]+\kappa_0D[a_0]\rho+\gamma_0D[a^{\dag}_0a_0]\rho\cr
&&\cr
&&+\sum_{k=1,2}\kappa_k D[a_k]\rho+\gamma_kD[a^{\dag}_k a_k]\rho,
\end{eqnarray}
where $\rho$ is the density operator of the system, $D[o]\rho=o\rho o^{\dag}-\frac{1}{2}(o^{\dag}o\rho+\rho o^{\dag}o)$. For simplicity, we assume that $\kappa_k=\kappa$ $(k=1,2)$ is the single-photon losses rate, and $\gamma_k=\gamma$ is the pure dephasing rate. Note that the influence of decoherence in the cavity mode $\mathcal{A}_0$ is different from those in the KPOs. The decoherence of the cavity mode $\mathcal{A}_0$ can be adiabatically eliminated for large $\Delta$  \cite{PhysRevApplied.18.024076}, so the system is insensitive to the decoherence of $\mathcal{A}_0$. In the numerical simulation, we set $\kappa_0=\kappa$ and $\gamma_0=\gamma$ for convenience. To well understand the effect of decoherence, we project the system onto the eigenstates of $H_k^{\mathrm{Kerr}}$. The projection operator of the KPOs is
\begin{eqnarray}\label{25}
P=\sum_{k=1,2}[|C_{\pm}\rangle_k\langle{C_{\pm}}|+\sum_{v=1}^{\infty}|\psi_{\pm}^{e,v}\rangle_k\langle\psi^{e,v}_{\pm}|],
\end{eqnarray}
where $|\psi^{e,v}_{\pm}\rangle_k=N^e_{\pm}[D_k(\alpha)\mp| D_k(-\alpha)]|v\rangle_k$, $v$ represents the level of the excited states.
\begin{figure}
  \centering
  \subfigure{\includegraphics[width=9cm]{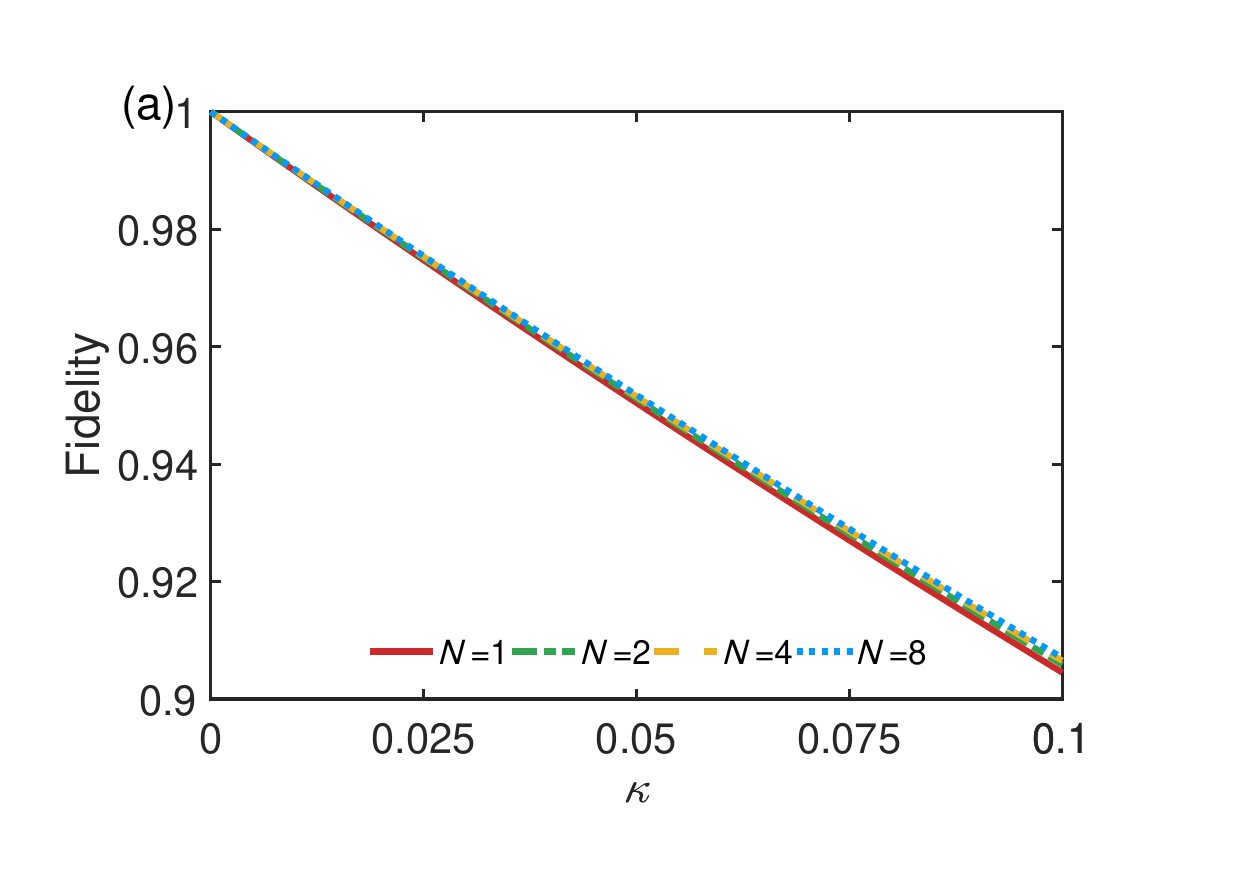}\label{k1248}}
  \subfigure{\includegraphics[width=9cm]{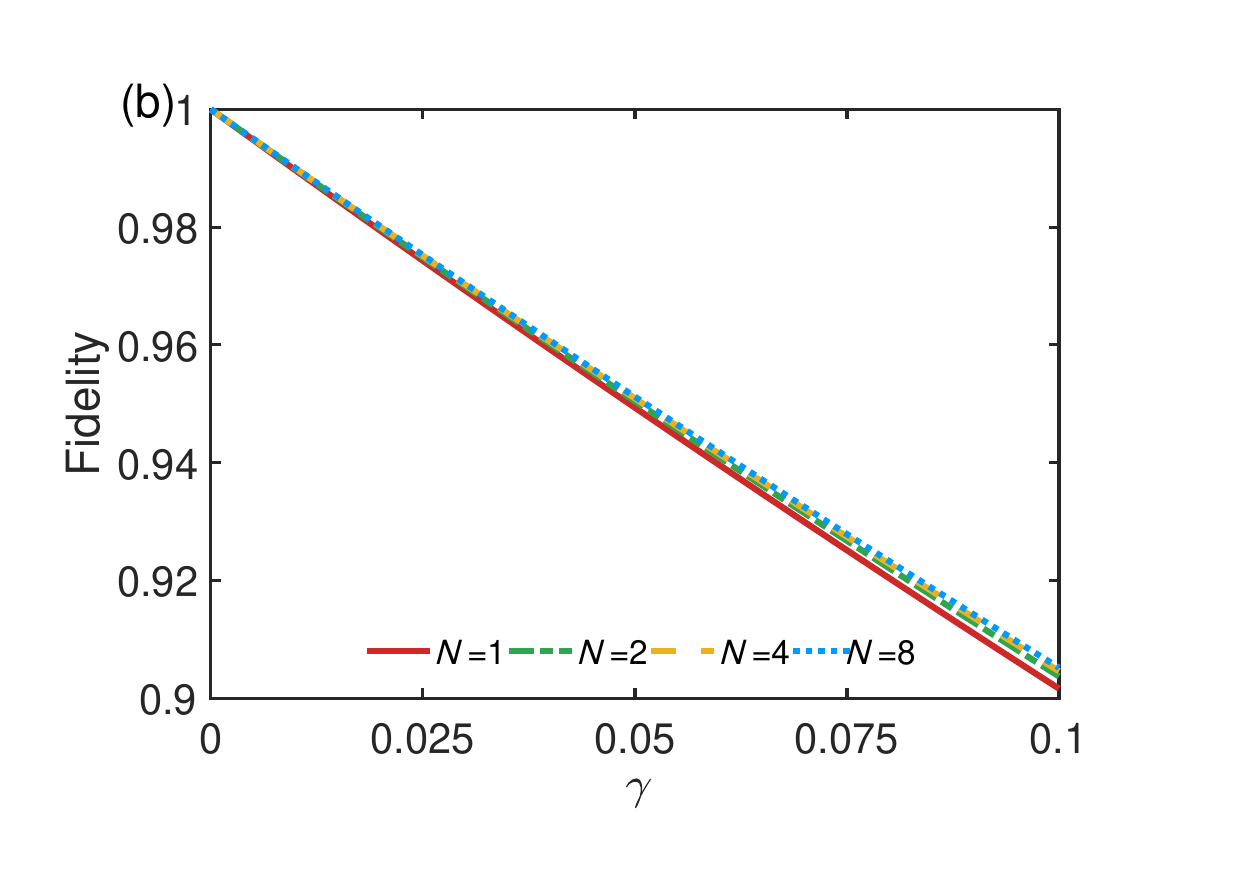}\label{r1248}}
  \caption{(a) Fidelities versus the single-photon losses rate $\kappa \in [0, 0.1]$ $\mathrm{MHz}$ and the dephasing rate $\gamma=0$. (b) Fidelities versus dephasing $\gamma \in [0,0.1]$ $\mathrm{MHz}$ with single-photon losses rate $\kappa=0$.}
\end{figure}
\begin{figure}
  \centering
  \includegraphics[width=8cm]{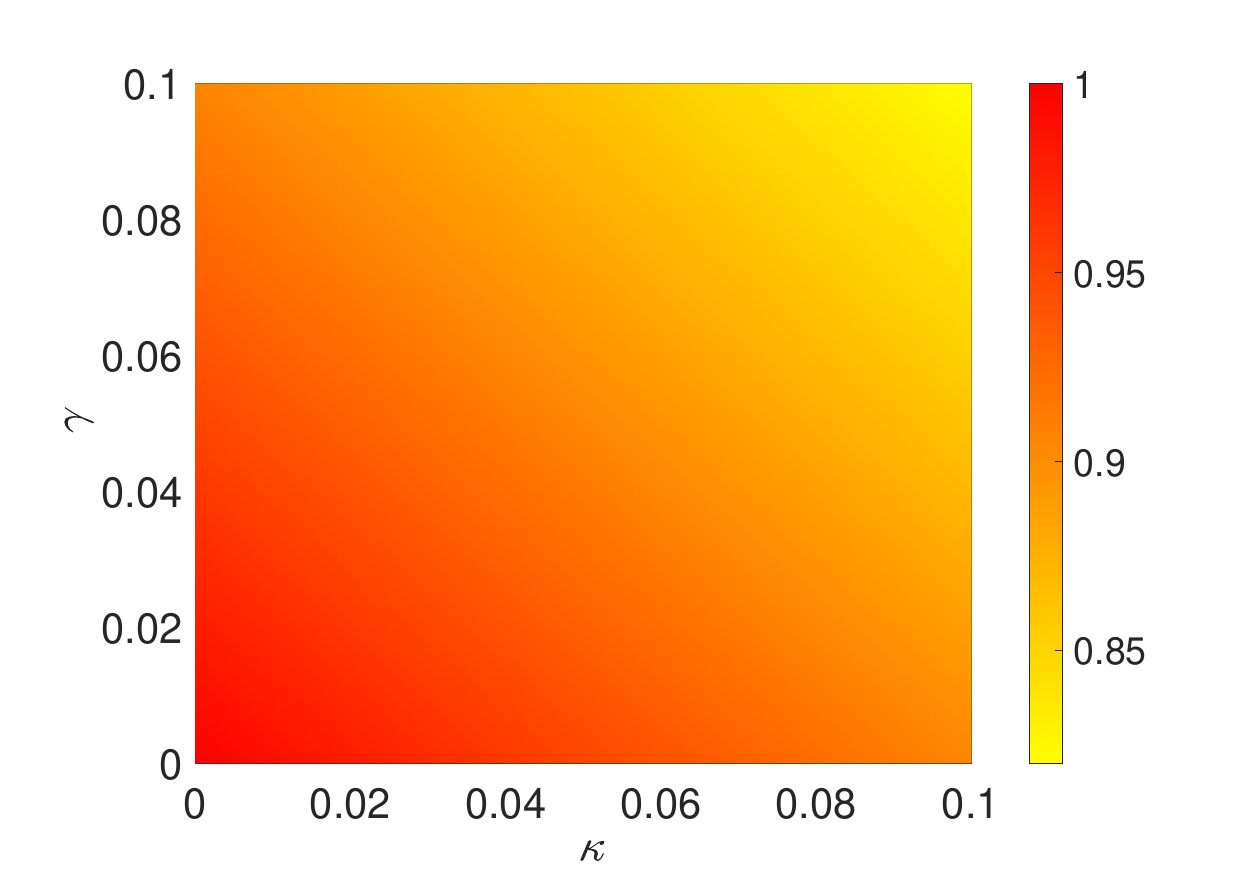}
  \caption{Fidelities of the optimized protocol with $N=8$ versus $\kappa=\gamma\in [0,0.1]$ $\mathrm{MHz}$. Other parameters remain unchanged.}
  \label{rk}
\end{figure}
Then the master equation becomes
\begin{eqnarray}\label{26}
\dot{\rho}&\simeq&-i[PHP,\rho]+\kappa_0D[a_0]\rho+\gamma_0D[a^{\dag}_0a_0]\rho\cr
&&\cr
&&+\sum_{k=1,2}\kappa_k D[Pa_kP]\rho+\sum_{k=1,2}\gamma_kD[Pa^{\dag}_k a_kP]\rho.
\end{eqnarray}
When $\gamma_k$, $\kappa_k$ are much smaller than the energy gap $E_{\mathrm{gap}}$, the dynamics of the system is still well confined to the cat-state subspace \cite{CAI202150,PhysRevX.9.041009,Puri2017,Puri2020}.
For large $\alpha$, the influence of the single-photon losses in the KPOs is described by the penultimate term in Eq.~(\ref{26})
\begin{eqnarray}\label{27}
D'[a_k]\!=&&D[Pa_kP]\cr\cr
\simeq&&\alpha^2D[\sqrt{\tanh\!\alpha^2}\!|C\!_+\rangle\!_k\!\langle C_-|\!+\!\sqrt{\coth\!\alpha^2}|C_-\rangle\!_k\!\langle\!C_+|]\cr\cr
&&+\!D[\sqrt{\frac{N_+}{N_+^e}}|C_+\rangle\!_k\!\langle\psi^{e,1}_+|
\!+\!\sqrt{\frac{N_-}{N_-^e}}|C_-\rangle\!_k\!\langle\psi^{e,1}_-|]\cr\cr
&&+\!\alpha^2D[\sqrt{\frac{N^e_-}{N_+^e}}|\psi^{e,1}_-\rangle\!_k\!\langle\psi^{e,1}_+|
\!+\!\sqrt{\frac{N_+^e}{N_-^e}}|\psi^{e,1}_+\rangle_k\langle\psi^{e,1}_-|],\cr\cr
&&
\end{eqnarray}
where $|\psi^{e,1}_{\pm}\rangle_k$ are the first-excited eigenstates of the Hamiltonian $H_k^{\mathrm{Kerr}}$.
Note that the highly excited eigenstates of KPOs are not excited in the presence of single-photon losses. Therefore, we only consider the first-excited eigenstates in Eq.~(\ref{25}). The second term in Eq.~(\ref{27}) means that single-photon losses can only cause the transition from the excited eigenstate $|\psi^{e,1}_{\pm}\rangle_k$ to the ground state $|C_{\pm}\rangle_k$. If the KPOs are initially in the cat-state subspace, they always remain in the cat-state subspace. Therefore, we can ignore the last two terms of the Eq.~(\ref{27}) and obtain
\begin{eqnarray}\label{28}
D'[a_k]\rho\simeq\frac{\alpha^2}{\sqrt{1-e^{-4\alpha^2}}}
D[\sigma^x_k+ie^{-2\alpha^2}\sigma_k^y]\rho,
\end{eqnarray}
where $\sigma^x_k=|C_+\rangle_k\langle C_-|+|C_-\rangle_k\langle C_+|$, $\sigma^y_k=i(|C_-\rangle_k\langle C_+|+|C_+\rangle_k\langle C_-|)$. Substitute Eq.~(\ref{28}) into Eq.~(\ref{26}), we find that in the computational subspace, single-photon losses mainly lead to a bit-flip error $\sigma^x_k$, accompanied by an exponentially small phase flip-error $\sigma^y_k$.

In Fig.~\ref{k1248}, we plot the fidelity of the single-photon losses for different values of $N$. The single-photon losses rate is assumed as $\kappa\in[0,0.1]$ $\mathrm{MHZ}$, and the dephasing rate is assumed as $\gamma=0$. As can be seen from Fig.~\ref{k1248}, the increase of $N$ does not affect the fidelity too much. The fidelities are all above 0.9 when $\kappa=0.1\mathrm{MHZ}$, which means that the optimized protocol still maintains a high fidelity in the presence of the single-photon losses. We also numerically simulate the dependence of the fidelity on the pure dephasing rate in Fig.~\ref{r1248} and a similar result is obtained.

Since single-photon losses and dephasing may coexist in the system, so we numerically investigate the effect on the fidelity when single-photon losses and dephasing are present together. {In Fig.~\ref{rk}, we plot the fidelities of the optimized protocol with $N=8$ versus $\kappa=\gamma\in[0,0.1]$ $\mathrm{MHz}$. As it is shown, when both types of noise are present, the entangled cat states can still maintain high fidelities for small noise rates. However, as $\kappa$ and $\gamma$ gradually increase, the fidelities will be significantly affected \cite{PhysRevApplied.18.024076}.}

\section{CONCLUSION}\label{V}

In conclusion, we have investigated a protocol to use photonic cat-state qubits for preparing high-fidelity entangled cat states. The entangled cat states can be successfully obtained under the action of the cat-code MS gate. We further adopt composite pulses to improve the fidelity of the entangled cat states. The modulated time-independent composite drives can be realized by introducing composite two-photon squeezing drives. Numerical simulation results show that the protocol can provide strong robustness against timing error and detuning error. Furthermore, under the influence of decoherence, the protocol still maintains a high fidelity. That is, stable and high-fidelity entangled cat states are possible with the large $N$. Meanwhile, using composite pulses to enhance the robustness of the protocol is feasible. We hope that the proposal could offer a simple method for preparing stable and high-fidelity entangled cat states in quantum computation.

\begin{acknowledgments}
Y.-H.C.
was supported by the National Natural Science Foundation of
China under Grant No. 12304390, Fujian 100 Talents Program, and Fujian Minjiang Scholar Program. Y. X. was supported by the National Natural Science Foundation of China under Grant No. 62471143, the Key Program of National Natural Science Foundation of Fujian Province under Grant No. 2024J02008, and the project from Fuzhou University under Grant No. JG2020001-2. 

{\begin{appendix}
\section{Magnus eapansion}\label{AA}
The Magnus expansion is a method used to find an approximate solution of the time-dependent Schr$\ddot{\mathrm{o}}$dinger equation \cite{Magnus1954OnTE,Blanes_2010,BLANES2009151,WANG2021127033}. Considering the Hamiltonian of the system as $H(t)$, the dynamics of the system is described by the Schr$\ddot{\mathrm{o}}$dinger equation in the interaction picture
\begin{eqnarray}\label{A1}
i\frac{d}{dt}|\psi(t)\rangle=H_I(t)|\psi(t)\rangle.
\end{eqnarray}
The time evolution of the system is determined by the evolution operator $U(t)$
\begin{eqnarray}\label{A2}
|\psi(t)\rangle=U(t)|\psi(0)\rangle,
\end{eqnarray}
where the evolution operator $U(t)$ satisfies the following differential equation
\begin{eqnarray}\label{A3}
i\frac{d}{dt}|U(t)\rangle=H_I(t)|U(t)\rangle.
\end{eqnarray}
The Magnus expansion relies on the assumption that there exists an exponential solution, which is given by \cite{PhysRevApplied.18.024076,WANG2021127033}
\begin{eqnarray}\label{A4}
U(t)=\exp[-i\sum^{\infty}_{l=1}\Omega_l(t)].
\end{eqnarray}
Under the large detuning condition \cite{WANG2021127033}, only the first two terms of $\sum^{\infty}_{l=1}\Omega_l(t)$ are retained. The expressions are as follows
\begin{eqnarray}\label{A5}
\Omega_1(t)&=&\int^{t}_{0}dt_1H_I(t_1),\cr\cr
\Omega_2(t)&=&\frac{1}{2}\int^{t}_{0}dt_1\int^{t_1}_{0}dt_2
\left[
\begin{array}{c}	
H_I(t_1),H_I(t_2)
\end{array}
\right].
\end{eqnarray}
Using the effective Hamiltonian in Eq.~(\ref{8}) to solve the Eq.~(\ref{A5}), we can obtain
\begin{eqnarray}\label{A6}
\Omega_1(t)&=&S_x(\chi(t)a_0^{\dagger}+\chi^{*}(t)a_0),\cr\cr
\Omega_2(t)&=&-\beta(t)S^2_x.
\end{eqnarray}
Correspondingly, the evolution operator of the system can be derived as
\begin{eqnarray}\label{A7}
U(t)=\exp\{-i(\chi(t)a_0^{\dagger}+\chi^{*}(t)a_0)S_x+\beta(t)S^2_x]\},
\end{eqnarray}
which corresponds to Eq.~(\ref{9}).
\end{appendix}}
\end{acknowledgments}
%

\end{document}